\documentstyle[aaspp4,eps,boxedeps]{article}

\def\kms{\,km\thinspace s$^{-1}$ }                  %kms -1%

\lefthead{Jerjen {\it et~al.}}
\righthead{THE SBF METHOD IN dE GALAXIES}

\begin{document}

\title{Testing the Surface Brightness Fluctuations Method for Dwarf 
Elliptical Galaxies in the Centaurus A Group}

\author{H.~Jerjen}
\affil{Research School of Astronomy and Astrophysics, Mt.~Stromlo and Siding Spring 
Observatories, ANU, Private Bag, Weston Creek PO, ACT 2611, Canberra, Australia \\
Electronic mail: jerjen@mso.anu.edu.au}

\author{K.C.~Freeman}
\affil{Research School of Astronomy and Astrophysics, Mt.~Stromlo and Siding Spring 
Observatories, ANU, Private Bag, Weston Creek PO, ACT 2611, Canberra, Australia \\
Electronic mail: kcf@mso.anu.edu.au}

\and

\author{B.~Binggeli}
\affil{Astronomical Institute of the University of Basel, Venusstrasse 7, CH-4102 
Binningen, Switzerland, \\ Electronic mail: binggeli@astro.unibas.ch}

\begin{abstract}
We have obtained deep $B$ and $R$-band CCD photometry for five 
dwarf elliptical galaxies that were previously identified on Schmidt
films covering the region of the Centaurus A (Cen\,A) group. From a 
Fourier analysis of the $R$-band CCD images we determined the surface 
brightness fluctuation (SBF) magnitude $\bar{m}_R$ for each stellar system. 
All magnitudes are similar and, given the small colour spread, 
suggest that these low surface brightness galaxies lie approximately at the 
same distance, regardless of the assumed SBF zero point. Long-slit spectra 
have been acquired to derive redshifts for two of the dwarfs, ESO269-066 and 
ESO384-016. The velocities, $v_\odot=784$\,\kms and $v_\odot=561$\,\kms, 
respectively, identify them unambiguously as Cen\,A group members. An age 
(H$\delta_{\mbox{A}}$) -- metallicity (C$_2 \lambda$4668) analysis of the 
spectra reveals an underlying old and metal-poor stellar population in both 
cases. Combining photometric and spectroscopic results we find strong evidence 
that indeed all dwarf galaxies are Cen\,A group members. 

Based on Cepheid, TRGB, and PNLF distances published for the two main 
Cen\,A group galaxies NGC5128 and NGC5253, we adopted a mean group distance
of 3.96\,Mpc to calibrate the apparent fluctuation magnitudes. The resulting 
{\em absolute}\/ SBF magnitudes $\bar{M}_R$ of the dEs correlate with the 
dereddened colours $(B-R)_0$ as predicted by Worthey's stellar synthesis models 
using the theoretical isochrones of Bertelli and collaborators. This good agreement 
allows a calibration of the SBF method for dwarf ellipticals in the colour range 
$0.8<(B-R)_0<1.5$. However, two branches of stellar populations appear in the 
$\bar{M}_R$-colour plane, and care has to be taken to decide which branch applies 
to a given observed dwarf. For dwarfs with $(B-R)_0<1$ there is very little colour 
dependence ($\bar{M}_R \approx -1.2$), in accord with our previous SBF analysis of faint,
blue Sculptor group dEs. For red dwarfs, $(B-R)_0>1.2$, the $\bar{M}_R$-colour 
relation is steep, and accurate colours are needed to achieve SBF distances
with an uncertainty of only 10\%. One of the dwarfs, ESO219-010, is located 
slightly behind the core of the Cen\,A group at $\sim$4.8\,Mpc, while the remaining four recover
the mean group distance of 3.96 Mpc that was put into the calibration. 
The depth of the group is only 0.5\,Mpc which identifies the Cen\,A group as 
a spatially well isolated galaxy aggregate, in contrast to the nearby Sculptor 
group.
\end{abstract}


\section{INTRODUCTION}
Dwarf elliptical (dE) galaxies are the most numerous type of galaxies 
in the nearby universe. Their existence is tightly correlated with the 
effective galaxy density of the environment, i.e.~dEs reside predominantly in
cluster cores (Binggeli et al.~1985; Ferguson 1989; Ferguson \& Sandage 
1990; Jerjen \& Dressler 1997) or accompany giant galaxies as satellites 
in the field (Binggeli, Tarenghi \& Sandage 1990). This naturally renders dEs 
as ideal objects to estimate distances to all sorts of galaxy 
aggregates. However, little advantage could be taken of the 
dE clustering properties to date, mainly because accurate distance
measurements to dEs still rely on the resolution of the galaxy light
into stars, such as any method based on the colour-magnitude 
diagram (CMD, e.g.~Smecker-Hane et al.~1994; Da\,Costa et al.~1996; 
Stetson 1997), or the tip of the red giant branch (TRGB, Harris et 
al.~1998; Grebel \& Guhathakurta 1999). This requirement makes observations 
expensive, restricting the practicability essentially to local systems. 

An alternative, very promising distance estimator for dEs is based on
``surface brightness fluctuations'' (SBF). The 
theoretical framework of the SBF method was introduced and discussed by 
Tonry \& Schneider (1988). The idea is to quantify the pixel-to-pixel brightness 
variation on a CCD image of a gas-free stellar system due to the varying 
numbers of stars within each pixel. At a constant surface brightness (star
density) this variation is then inversely proportional to the distance of the 
galaxy. In practice, one has to measure the so-called apparent fluctuation 
magnitude $\bar{m}_X$ in the photometric passband $X$. This quantity 
is the ratio of the second and first moments of the underlying stellar 
luminosity function. If sufficient information is known about the 
stellar content of a galaxy type, the absolute fluctuation magnitude $\bar{M}_X$
can be predicted from stellar synthesis models or empirically 
determined from calibrators, thus yielding a distance modulus of the galaxy.

So far SBF applications concentrated on high-surface brightness giant 
ellipticals (Tonry, Ajhar \& Luppino 1989, 1990; Tonry \& Schechter 1990; 
Pahre \& Mould 1994; Thomson et al.~1997; Tonry et al.~1997), bulges of 
spiral galaxies (Tonry 1991; Luppino \& Tonry 1993), and globular 
clusters (Ajhar \& Tonry 1994). The first attempt to obtain SBF distances 
for {\em dwarf}\/ ellipticals was made by Bothun et 
al.~(1991), followed only 
recently by our study of nearby dEs (Jerjen, Freeman, \& Binggeli~1998, hereafter 
JFB98) located in the outskirts of the Local Group (LG). In the latter study, 
the $R$-band SBF distances for five dEs in the Sculptor group were measured 
to an internal accuracy of 5--10\%. The difficulty with the SBF method for
dwarfs is the inherent difference in the stellar populations of 
giant and dwarf ellipticals, preventing the adoption of existing calibration
values from previous work on giants. To bypass the lack of empirical results, 
we based the SBF distances on the zero point predictions of stellar synthesis 
models (Worthey 1994). As it turned out, the calibrating $\bar{M}_R$ is fairly
insensitive to the star formation history for these very faint and blue
Sculptor dEs. The resulting distances showed that the dwarf galaxies 
closely follow the spatial 
distribution of the main Sculptor group galaxies, which are spread over a 
distance range between 1.7\,Mpc and 4.4\,Mpc. 

However, this large line-of-sight 
extension of the Sculptor group is clearly not ideal for the empirical calibration
of the SBF method. To improve this situation, we continue our SBF project 
here with a study of five dE galaxies identified in Centaurus\,A (Cen\, A) 
group region. There are two major advantages of working in 
the Cen\,A group rather than the Sculptor group. Firstly, the Cen\,A group is physically 
and spatially much better defined than the Sculptor group. The Sculptor aggregate is,
in fact, an unbound and almost freely expanding ``cloud'' of galaxies (see JFB98),
while Cen\,A is certainly a rich and well concentrated group with a small distance
dispersion of the member galaxies. Secondly, the Cen\,A group's mean 
distance is well established via Cepheids and tip of the red giant branch 
magnitudes measured for the two main group galaxies, NGC5128 (= Centaurus\,A) 
and NGC5253. 

Owing to their greater mean distances, the Cen\,A group dwarf Es are also 
intrinsically brighter on average than the Sculptor dwarfs. Thus in this paper we also 
extend the calibration of the SBF method to brighter magnitudes. As it turns out, 
for brighter dEs the calibration is somewhat more difficult. Brighter dwarfs are on
average also {\em redder}\/ than faint ones, and for red objects, according to the 
stellar synthesis models, the calibrating $\bar{M}_R$ depends rather strongly 
on the colour. Moreover, in a certain colour range the calibration is not unique. 
However, we show that by a proper treatment of the $\bar{M}_R$ - colour relation(s) 
and with reliable colour data, accurate SBF distances can be derived also for brighter dEs.

The paper is organized as follows. In \S 2 we describe the observations and
data reduction procedures. This includes the spectroscopy of two of
the five sample dEs, for which radial velocities as well as ages and metallicities 
are derived. \S 3 is the SBF analysis resulting in the apparent fluctuation
magnitudes of the galaxies. In \S 4 these fluctuation magnitudes are then
calibrated by means of stellar synthesis models which are critically discussed
in terms of the $\bar{M}_R$ - colour dependence. Summary and conclusions
are given in \S 5.

\section{OBSERVATIONS AND DATA REDUCTION}
As reported elsewhere (Jerjen, Binggeli, \& Freeman 2000, hereafter JBF00), 
13 low surface brightness dwarf galaxies with dE morphology were identified 
in the region of the Cen\,A group. They have been detected as part of a  
visual inspection of $\approx 50$ fine-grain IIIa-J emulsion SRC Schmidt films 
covering the group's area: $12^h 30^m < \mbox{R.A.(1950)} < 15^h,\, -20^\circ 
< \mbox{Decl.(1950)} < -50^\circ$. For the present SBF study we selected a subsample 
of five dEs. These galaxies have sufficiently large isophotal diameters providing 
enough independent measure points for the SBF analysis and their profiles are 
mostly unaffected by the disturbing light of bright nearby stars. In Figure~\ref{fig9}  
we show the distribution of the sample dEs on the sky.

\subsection{Imaging}
The photometric data for the five dwarf galaxies were taken with the 2.3m ANU 
telescope at Siding Spring Observatory in the dark time period 1997 April 9--11. 
The CCD camera was mounted at the Nasmyth focus B. On the first night the detector 
was a 2k$\times$1k EEV thick CCD with 22$\mu$m pixels and a pixel scale of 
0.55$''$. On the remaining two nights we used a 2k$\times$800 SITe thinned CCD, 
AR coated, 15$\mu$m pixels, and a pixel scale of 0.375$''$. The CCD gain 
was set to 1$e^-$ ADU$^{-1}$ for all exposures. The field of view is mechanically 
restricted to a $6'.7$ diameter circle. 

All three nights were photometric and seeing ranged from 1.1 to 1.4$''$ FWHM. 
A series of 4--5 $R$-band images of 600--720\,sec duration were taken for each galaxy 
yielding a total exposure time of 2400--3600\,sec per object. Such long integration 
times are required to allow the SBF signal from a low-surface brightness galaxy to 
grow and surpass the high level of sky shot noise present in the power spectrum. 
This noise level continuously increases when working with redder photometric passbands 
and is rather high in $I$ and $K$, the favourite filters for SBF studies on giant 
ellipticals. Even though the fluctuation signal also gets stronger this only partially 
compensates for the brighter sky, thus the Cousins $R$-band appears to be the best 
compromise for the work on dwarf ellipticals. Furthermore, with $R$-band images we 
avoid fringing effects which occur with thinned CCDs beyond 7000\AA\, and which can 
severely affect the weak fluctuation signal. 

The $R$-band observations of a galaxy were supplemented with a series of $6\times$600sec 
$B$ band images to get colour information. Besides the science images we took high 
signal-to-noise twilight flats every dusk and dawn period and photometric standards 
in the E regions (Graham 1982) throughout the nights.

The CCD data were reduced with standard IRAF procedures. The bias was subtracted 
from all of the images, and we used the combined twilight flats to flatten the science images. 
On each frame we determined the sky level from 10--15 selected star-free regions well away 
from the galaxies. The rms scatter showed that all images were flat to $0.1-0.2$\% over the 
full frame. Each set of sky-subtracted images of a galaxy was registered and median combined 
to generate a composite image. The calibration of these master images was carried out with 
aperture photometry of the standard stars. Uncertainty in the photometric zero point was 
measured at 0.02 mag.  

\begin{figure}
\centering\leavevmode
\epsfxsize=6.5cm
\framebox[6.6cm][c]{\epsfbox{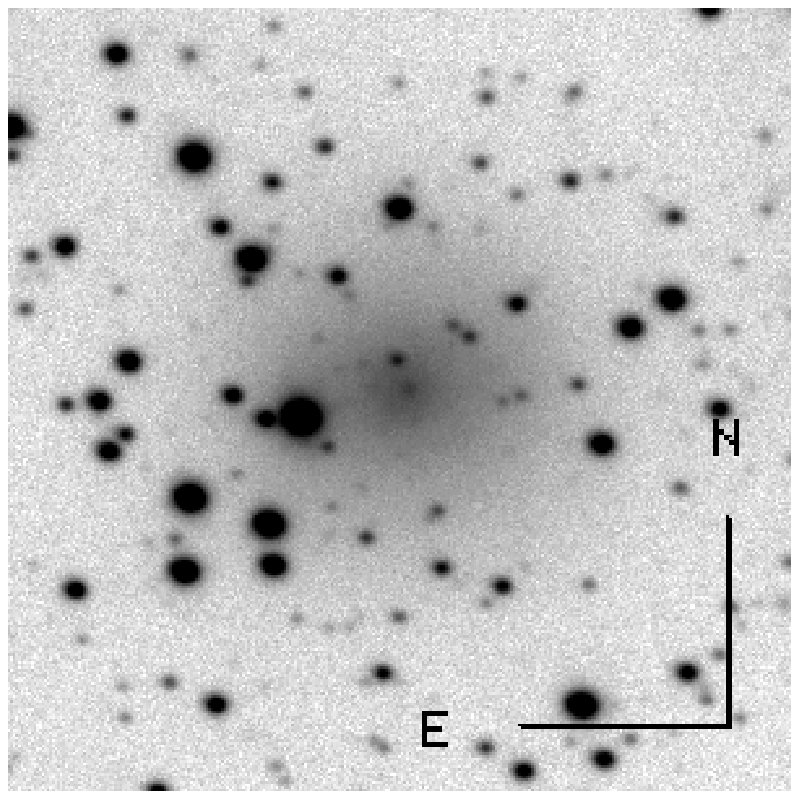}}
\caption{Deep $R$-band CCD image of ESO219-010. 
The length of the orientation bars is 30 arcsec.\label{fig1}}
\end{figure}

\begin{figure*}
\centering\leavevmode
\epsfxsize=6.5cm
\framebox[6.6cm][c]{\epsfbox{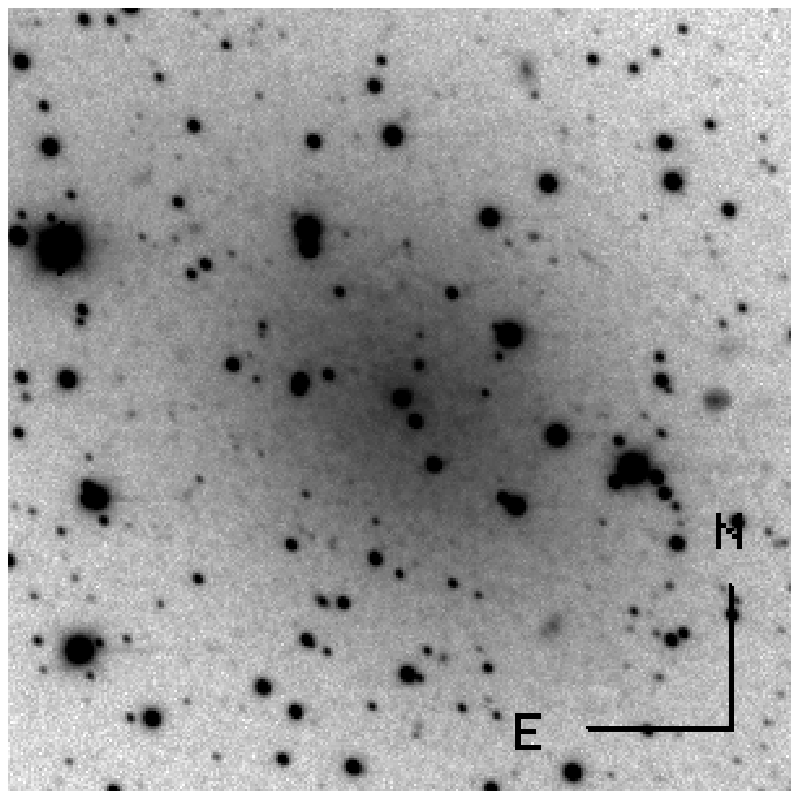}}
\epsfxsize=6.5cm
\framebox[6.6cm][c]{\epsfbox{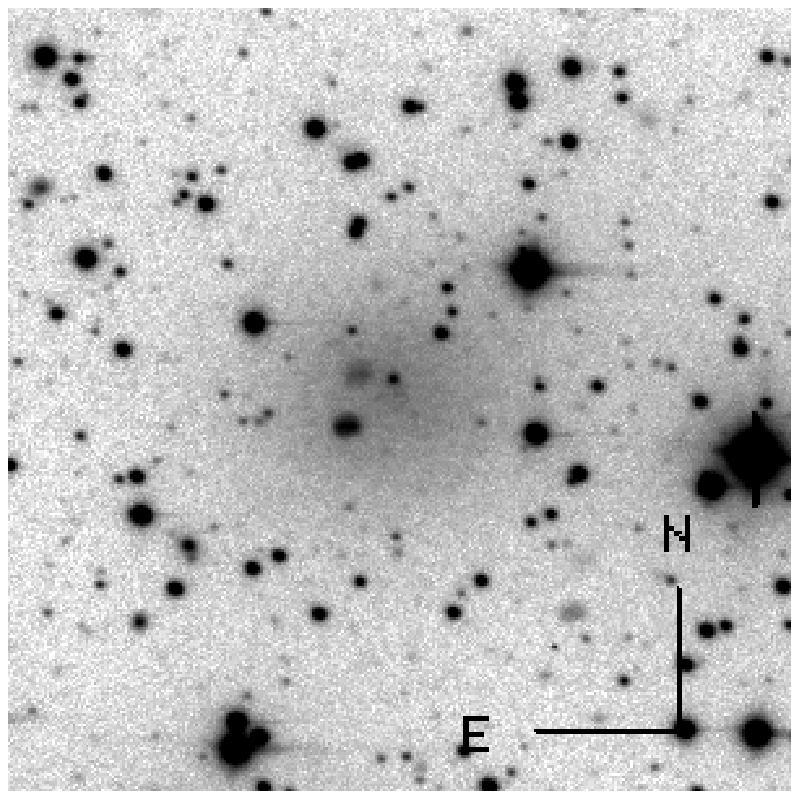}}\\

\centering\leavevmode
\epsfxsize=6.5cm
\framebox[6.6cm][c]{\epsfbox{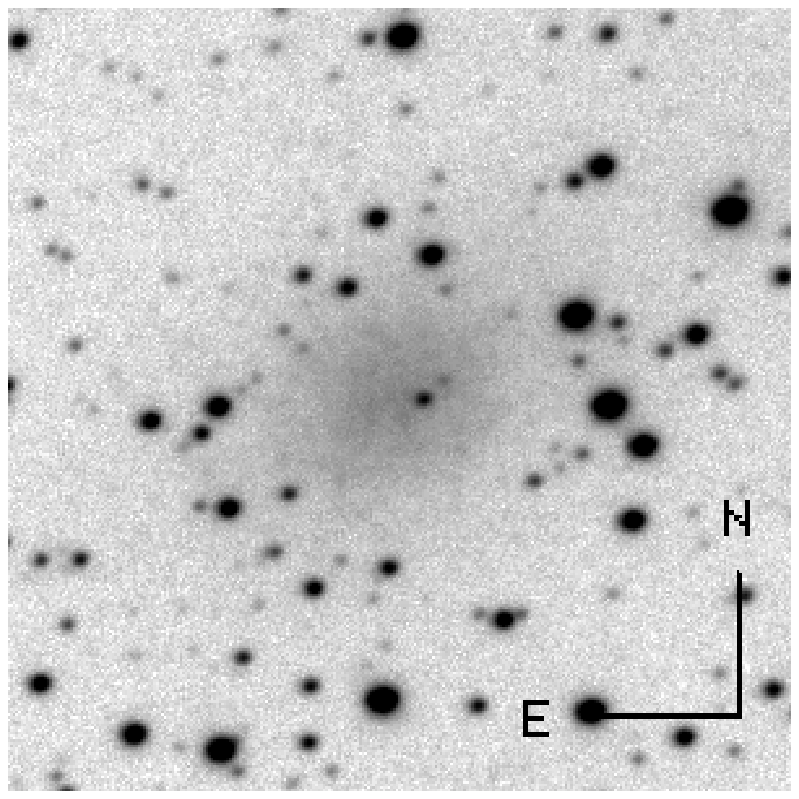}}
\epsfxsize=6.5cm
\framebox[6.6cm][c]{\epsfbox{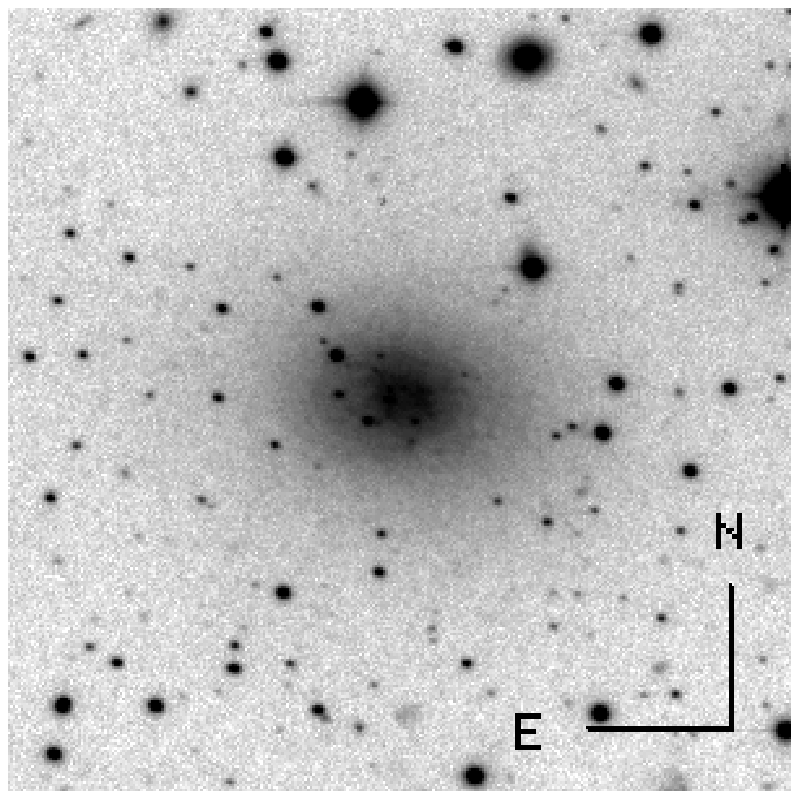}}
\caption{Deep $R$-band CCD images of ESO269-066, 
AM1339-445, AM1343-452, and ESO384-016 (l-to-r and t-to-b). The length 
of the orientation bars is 30 arcsec.\label{fig2}} 
\end{figure*}

Figures \ref{fig1} and \ref{fig2} show the $R$-band images of the five dEs. 
Among early-type dwarf galaxies there is some morphological variety (see Sandage 
\& Binggeli 1984). At the bright end of the dE luminosity function, there is a variant 
that was named dS0, because it is often distinguished by a S0-like, two-component 
structure (see also Binggeli \& Cameron 1991). ESO384-016 is clearly of this type, 
ESO269-066 arguably so. Bright dwarf ellipticals also host quite often an unresolved, 
centrally located star-like object, which is possibly a massive globular cluster formed 
in or fallen into the core region of the galaxy. ESO219-010 and ESO269-066 are the 
galaxies that show qualitative evidence for such a nucleus. However, a star projected 
onto the galaxy centre by chance cannot be ruled out in either case; thus a ``:'' goes 
with the classifications. 

In Table~\ref{tbl-1} we list the fundamental parameters of the dwarf galaxies: the 
morphological type (col.~[2]), coordinates (cols.[3-4]), the total $B$ and $R$ 
magnitudes (cols.~[5-6]), the effective radius $r_{eff,R}$ containing half of 
the total light (col.~[7]), and the mean effective surface brightness $\langle 
\mu \rangle_{eff,R}$ (col.~[8]). A full account of the observations including a 
discussion of the radial surface brightness profiles, colour gradients, and 
structure parameters of the galaxies is given in JBF00. The dereddened 
overall $(B-R)^T$ colours (col.~[10]) were derived from cols.[5-6] and corrected for 
foreground extinction using the IRAS/DIRBE maps of dust IR emission 
(Schlegel et al.~1998) and the ratio $A_B: A_R : E(B-V)=4.315: 2.673: 1$. 
According to these authors the extinction estimates (col.~[9]) have an accuracy 
of $16$\%. The observed colour range $1.1< (B-R)_0^T < 1.5$ (col.~[10]) covered by 
our Cen\,A group candidates is redder than for the Sculptor dEs (JFB98) 
and again is typical for dEs whose $B - R$ values range from 0.5 to 2.3 (Evans et 
al.~1990). 

\subsection{Spectroscopy}
The two sample galaxies with the highest surface brightness were selected for a 
spectroscopic follow-up to estimate their ages and metal abundances and to 
measure their redshifts. We obtained two spectra of ESO269-066, on April 
19, 1996 and two spectra of ESO384-016 on February 2, 1997, at the Nasmyth 
A focus of the 2.3m ANU telescope using a 600 line grating at the blue side of 
the double-beam spectrograph. Each exposure was of 2000\,sec duration. The 
detector was a SITe 1752$\times$532 thinned CCD with an across dispersion 
of $0.91''$\,pixel$^{-1}$ and a spectral resolution of 1.1\AA\, pixel$^{-1}$. 
The grating angle was set to cover the wavelength range $3500$\AA$-5500$\AA. We 
acquired the data using a long slit ($2''\times 6'.7$) under good seeing 
conditions. The slit was positioned at the galaxy centre and aligned along 
the major axis. 

\begin{figure}
\centering\leavevmode
\epsfxsize=10cm
\epsfbox{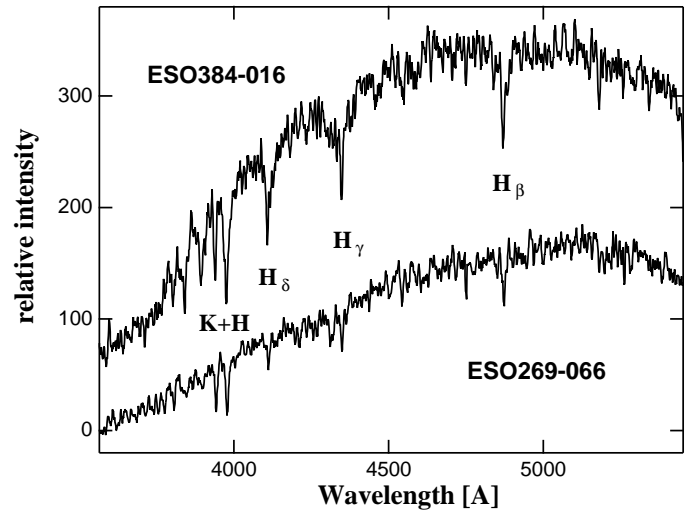}
\caption{Absorption line spectra of ESO384-016 and ESO269-066.
The prominent Balmer lines and the Calcium II H and K doublet are marked.
\label{fig3}}
\end{figure}

We reduced the CCD data in a straightforward manner with standard IRAF 
procedures. After the bias level was removed, each image was flatfielded 
and sky line subtracted. Wavelength calibration was performed using emission 
line spectra of a Ne-Ar lamp observed immediately after each galaxy 
spectrum. The combined galaxy spectra are plotted in Figure~\ref{fig3}. 
Both dwarfs are pure absorption line systems with prominent Balmer 
lines (H$_\beta$ $\lambda 4861$, H$_\gamma$ $\lambda 4340$, H$_\delta$ 
$\lambda 4101$) and a \ion{Ca}{2} H and K doublet ($\lambda 3968$, 
$\lambda 3934$). All lines are slightly stronger in the spectrum of ESO384-016.

The age and metallicity of the dwarf galaxies were estimated (L. A. Jones 1999,
private communication) by 
comparing the observed index strengths of the lines C$_2\lambda$4668 (Worthey 1994) and 
H$\delta_{\mbox{A}}$ at $\lambda 4102.9$ (Jones \& Worthey 1995) with the index 
strengths computed from single-burst population models. The C$_2\lambda$4668 line 
is a blend from C$_2$, Mg, Fe, and other elements and is considered as a very good 
metallicity indicator with a 2--3 times better orthogonal separation
of age and metallicity than the most commonly used Mg$_2$ index (Jones \& 
Worthey 1995). On the other hand, H$\delta_{\mbox{A}}$ is relatively 
metallicity-insensitive, thus useful in determining age. 
Furthermore, H$\delta_{\mbox{A}}$ is 
less affected by emission from ionised hydrogen than for instance H$\beta$ and less 
diluted by light from giant stars. 

In Figure~\ref{fig4} we show the location of the 
dwarf elliptical galaxies in the H$\delta_{\mbox{A}}$ versus C$_2$4668 (age--metallicity) 
diagram. Both galaxies appear in the old, metal-poor part of the diagram where 
crowding in the parameter space leads to poor resolution. ESO384-016 seems to be 
slightly more metal-rich and younger relative to ESO269-066. However, a 
quantification is impossible due to large uncertainties. Also, it has to be borne in mind
that the assumed single-burst model need not be correct.

To determine the redshifts of the dwarfs, their spectra were continuum-subtracted 
and cross-correlated with the spectrum of the velocity standard 
HD176047, a K1III giant. A heliocentric velocity of v$_\odot=561(\pm32)$\kms 
was measured for ESO384-016, which is in good agreement with velocities 
observed for well-known Cen\,A group galaxies such as NGC5128 (v$_\odot$=562\kms) 
or NGC5236 (v$_\odot$=516\kms). For ESO269-066 we measured v$_\odot$=784($\pm$31) 
which lies at the high end of the velocity distribution of the Cen\,A group. The 
most complete catalogues of Cen\,A group galaxies (C\^ot\'e et al.~1997, hereafter 
C97; Banks et al.~1999) list three galaxies with comparable velocities: 
SGC1259.6-1659 (v$_\odot$=732\kms), DDO161 (v$_\odot$=747\kms), and AM1331-451 
(v$_\odot$=831\kms). A fourth galaxy, AM1318-444 ($13^h$ $21^m$ $47^s\!.1$, 
$-45^\circ$ $03'$ $49''$ J2000.0) with v$_\odot$=741\kms, was reported just recently 
(JBF00).  Assuming D=4\,Mpc, the small distance and velocity differences between ESO269-066 
and AM1331-451 ($250$\,kpc, $\Delta$\,v=47\kms) or AM1318-444 ($100$\,kpc, 
$\Delta$\,v=$-43$\kms) may indicate a small subclump within the group (see Figure~\ref{fig9}). 

Overall, the observed redshifts provide clear evidence that ESO269-066 and 
ESO384-016 are indeed Cen\,A group members. This picture is further qualitatively 
supported by the facts that (i) the Cen\,A group is well isolated in velocity space 
(Tully \& Fisher 1987; C97) i.e.~has no overlap in velocity space with a nearby background 
galaxy aggregate, and that (ii) dwarf ellipticals found in low density regions are 
almost exclusively close companions of massive galaxies (Binggeli, Tarenghi \& Sandage 1990; 
Ferguson \& Binggeli 1994; JFB98). A quantitative confirmation of these kinematic
distance estimates is of course provided by the following SBF analysis. 

\section{SBF ANALYSIS}
The deep $R$-band CCD master images of our galaxies were prepared according to the 
procedure 
described in TS88 and JFB98. Point sources were identified with the DAOPHOT routines 
(Stetson 1987) and the affected CCD areas were replaced by nearby uncontaminated 
patches of the galaxy from the same surface brightness range. The isophotes of 
the cleaned galaxy were 
modelled with the ISOPHOT package within STSDAS by fitting ellipses with variable 
radius, ellipticity, and position angle. The mean galaxy brightness distribution was 
subtracted from the master image. The residual image was then divided by the square root 
of the mean profile to produce the fluctuation image. This last step normalizes the 
strength of the pixel-to-pixel variation over the galaxy's surface due to differences
in surface brightness (stellar density).

On each fluctuation image, two subimages (F1 and F2) were selected within the 
25.5 $R$\,mag\,arcsec$^{-2}$ isophote of the galaxy for the SBF analysis. The only
exception was ESO219-010 where superimposed foreground stars and the relative small size 
of the galaxy allowed us to work with only one subimage. The size of a subframe was 
between $40\times 40$ and $70\times 70$ pixels. Under the given seeing
conditions this area corresponds to 900--4000 independent points carrying the SBF 
fluctuation signal. The overlap of subimages was kept minimal ($<5$\%) to get two 
independent distance measurements for a galaxy. Objects previously identified with 
DAOPHOT on a subimage were replaced by a randomly selected area from the region outside 
of the subimage but in the same isophotal range. The fraction of pixels patched in 
this way was less than 5\% of the total subimage area. All cleaned fluctuation images 
were Fourier-transformed and the azimuthally averaged power spectrum calculated 
(Figures~\ref{fig5} and \ref{fig6}). From well isolated bright stars on the master 
images we derived the power spectrum of the PSF. We then fitted a linear combination 
of the power spectrum of the PSF and a constant to the the galaxy power spectrum, 
\begin{equation} 
\mbox{PS}_{\mbox{gal}}(k)=P_0\cdot \mbox{PS}_{\mbox{PSF}}(k) + P_1.
\end{equation}

Low spatial frequencies ($k<4$) which can be affected by the galaxy subtraction 
were excluded from the fit. The results are listed in Table~\ref{tbl-2}. From the 
amplitude $P_0$ (col.~[7]) we calculated the apparent fluctuation magnitude 
(col.~[9]) with the formula $\bar{m}_R=m_1-2.5\cdot\log(P_0/t)$. Thereby, the quantity 
$m_1$ (col.~[2]) is the magnitude of a star yielding 1 ADU per second on the CCD 
and $t$ (col.~[3]) is the integration time of a single exposure. $P_1$ (col.~[8]) 
is the scale-free (white-noise) component in the power spectrum which is proportional 
to the ratio of the sky (col.~[6]) and the galaxy surface brightness (col.~[5]) averaged
over the subimage. To keep the white-noise level low it is thus crucial to restrict 
the SBF analysis to the high-surface brightness parts of the galaxy. The combination 
of $n$ exposures reduced the white-noise level by a factor of $\sqrt{n}$.

The overall error of the apparent fluctuation magnitude $\bar{m}_R^0$  is 
dominated by the power spectrum fitting error which contributes 3--10\%. 
Applying the SBF method to subimages of relatively high surface brightness 
($\bar{g}$, col.~[5]) helped to keep the error which is due to the uncertainty 
in the sky determination (col.~[6]) small, i.e.~3\% or less. Errors from 
PSF normalisation and the shape variation of the stellar 
PSF over the CCD area are equally small (1--3\%). In Column 7 of 
Table~\ref{tbl-2} we list the $P_0$ values with the combined error from all 
these sources. Assuming a photometric accuracy of $\Delta m_1=0.02$\,mag 
(col.~[2]) and adopting a 16\% error for foreground extinction (Schlegel 
et al.~1997), the formal internal uncertainty in $\bar{m}_R^0$ is between 
0.05 to 0.14\,mag (col.~[11]). We like to stress that the two narrow but 
distinct magnitude ranges covered by $m_1$ are due to the use of two 
different CCD detectors with different zero points (see section 2.1).

\section{CALIBRATION OF FLUCTUATION MAGNITUDES AND DISTANCES}
\subsection{Theoretical stellar populations}
First let us recall that most faint dEs are not single-burst populations like 
globular clusters (Da Costa 1997). A rather diverse and complex set 
of star formation histories (SFH) is observed among the local dwarf 
spheroidals. Their stellar populations range from old (Ursa Minor) 
and mainly old (e.g.~Tucana, Leo\,II) through intermediate-age episodic 
SFH (e.g.~Carina, Leo\,I) to intermediate-age continuous SFH (e.g.~Fornax). 
Phoenix and LSG3 are classified as dE/Im, because they show similarities 
to both dwarf spheroidals and dwarf irregulars. These systems are dominated 
by an old metal-poor population with no evidence for {\em major} star 
formation activities after the initial episode 8-10 Gyr ago. However, 
both systems have a minor population of young stars, with ages of about 
150 Myr, which makes these galaxies resemble dwarf irregulars. 

To explore the age- and metallicity-dependency of $\bar{M}_R$ for dE-like stellar 
populations we employed Worthey's on-line model interpolation 
engine \footnote{http://199.120.161.183:80/$\sim$worthey/dial/dial\_a\_model.html}.
We considered separately the two offered sets of isochrones: (1) the 
amalgamation of the stellar evolutionary isochrones of VandenBerg and 
collaborators and the Revised Yale Isochrones (Green et al.~1987; hereafter RYI) 
as described in Worthey (1994), and (2) the Padova isochrone library 
(Bertelli et al.~1994; hereafter PI).

We computed $\bar{M}_R$ and $(B-R)$ values for the following 2-component stellar 
populations. The age of the main population was set to 17, 12, or 8 Gyr and its 
metallicity set in the range $-2.0$\footnote{The model limits for the Bertelli 
isochrones are $-1.3$ for 17\,Gyr and $-1.7$ for 12 and 8\,Gyrs}, $-1.9$, ...,$-1.0$, 
$-0.5$, $-0.25$, $0$. The second population, 5 Gyr old 
and of solar metallicity, contributed in weight to the whole population at the 0 ($\sim$ 
pure case), 10, 20 or 30\% level. In all cases we assumed a Salpeter IMF. The 
results of the models are illustrated in Figure~\ref{fig7} where we have plotted 
$\bar{M}_R$ versus $(B-R)$. The models are separated according to the underlying 
isochrone library. In each case, two distinct branches are visible. However, only 
one branch is well defined, while the other, though recognisable, is only sparsely 
populated. We will call the steep linear and shallow quadratic component the red 
and blue branch, respectively,  for reasons given below. 

In order to understand the significant differences between the predictions of the 
two sets of isochrones for $(B-R)<1.3$ one has to recall that $\bar{M}_R$ 
corresponds to the luminosity-weighted luminosity of the underlying stellar 
population. This quantity is roughly the luminosity of a giant star, thus 
critically depends on the post red giant branch (RGB) evolution. While the 
theoretical isochrones of Bertelli et al.~(1994) are the most complete set available 
with a full RGB evolution implemented, this part of the stellar evolution is missing 
in the RYI. Worthey assumed in his RYI-based models the horizontal branch to 
remain in a red clump near the giant branch, which is incorrect for metal-poor 
populations. Because of these limitations we will focus in the following on the
results of the PI based models.
 
A few words on the interpretation of the PI colour-fluctuation magnitude
diagram. Da Costa (1997) pointed out that (i) the colour of the red giant branch 
is largely independent of age, but is strongly dependent on metallicity (his Figure\,1)
and (ii) that for a fixed metallicity the colour of the horizontal branch 
has a discontinuity from blue for the oldest populations ($\sim 15$\,Gyr) to red 
at younger ages ($\leq 10$\,Gyr, his Figure\,4). The latter effect is more marked for
lower metallicities. These two trends describe very accurately the behaviour of 
$\bar{M}_R$ above and below $(B-R)=-1.3$ (or [Fe/H]$\sim-1.0$) as we illustrate
in Figure~\ref{fig8}. In the models, only the very old (17\,Gyr) and metal-poor 
([Fe/H]$<-1.0$) populations are located on the blue branch. With an increasing
fraction of younger stars the total population gets redder and the model values 
reach quickly the red branch.  
 
To quantify the location of the two branches, we deduced two analytical expressions 
for the locus of theoretical values by least-squares fitting to the model data: 
$\bar{M}_R=6.09\cdot (B-R)_0-8.78$ for the red, and $\bar{M}_R=1.89\cdot 
[(B-R)_0-0.77]^2-1.23$ for the blue branch, respectively. 

\subsection{Observed stellar populations}
The observed $\bar{m}_R^0$ magnitudes (col.~[11], Table~\ref{tbl-2}) occupy 
a small range with $\sigma=0.3$\,mag which, given the small colour range,
implies roughly the same distance 
for our sample galaxies independently from any assumption on the zero point 
of the SBF distance estimator. Furthermore, we found quantitative evidence 
in \S 2.2 that two of the dEs are Cen\,A group members according to 
their redshifts. This immediately suggests that indeed {\em all} sample 
dEs are Cen\,A group galaxies. Based on this assumption we can investigate 
the empirical calibration of the SBF method for dEs. 

In Table~\ref{tbl-4} we compiled accurate distances of two main Cen\,A 
group galaxies from the literature. Three distances are published for 
NGC5128: Harris et al.~(1999) measured  $D=3.9(\pm0.3)$\,Mpc 
based on the tip of the red giant branch (TRGB) magnitude. 
A distance from the planetary nebulae luminosity function (PNLF) of 
$D=3.9(\pm0.3)$\,Mpc was given by Hui et al.~(1993), and $D=3.6(\pm 0.2)$\,Mpc 
was obtained by Tonry \& Schechter (1990) with the SBF method for giant 
ellipticals. The two latter values have been adjusted to the
presently favoured LG 
distance scale set by $(m - M)_0$(M31) = 24.5 (van den Bergh 1995; Fernley et 
al.~1998; Harris 1999). A mean value of $D=3.8(\pm0.1)$\,Mpc 
[(m--M)=$27.90\pm 0.05$] is taken as the true distance of NGC5128. The second 
galaxy is NGC5253. Saha et al.~(1995) reported a distance 
modulus of (m--M)$=28.08(\pm 0.10)$ derived from Cepheids. NGC5128 
and NGC5253 are the only Cen\,A group galaxies with accurate distances 
to date, which makes the estimation of the group distance difficult in principal. 
However, their redshifts (col.~[2], Table~\ref{tbl-4}) are close to the mean 
group velocity at v$_\odot$=551\kms (C97) and we adopt their average distance 
modulus (m--M)$=27.99(\pm0.17)$, or $D$ = 3.96 Mpc, for the whole Cen\,A group.
 
We note that more distance determinations are available for these two 
galaxies (e.g.~Della Valle \& Melnick 1992; Shopbell, Bland-Hawthorn \& Malin 1993; 
Soria et al.~1996) as well as for other group members (e.g.~De Vaucouleurs 
1979). However, these results have been superseded by newer data, or are 
less secure. 

Using the independent Cen\,A group distance we converted the fluctuation 
magnitude of each SBF field (col.~[4], Table~\ref{tbl-3}) into an absolute 
magnitude. The results are shown in Figure~\ref{fig7} whereby the 
colour (col.~[3], Table~\ref{tbl-3}) was determined for each SBF field 
individually because integrated colours as listed in col.~[10] of Table~\ref{tbl-1} 
are not reliable due to observed colour gradients (see JBF00). The estimated 
error in the local $(B-R)_0$ is composed of the sky uncertainty in the $B$ and 
$R$-band images, the photometric error, and an extinction error of 16\% 
(Schlegel et al.~1998).

\subsection{Theory vs.~observations and the final calibration}
A comparison between observations and theoretical models based on PI 
(Figure~\ref{fig7}, {\em right panel}) reveals a remarkable agreement. All dEs 
(except ESO219-010 which will be discussed separately below) are close 
to the locus of theoretical values. Each branch exhibited by the models 
is populated with the data of at 
least one sample galaxy. This confirms the previous assumption that 
the dEs are Cen\,A group members, and that the value of the adopted 
Cen\,A group distance modulus of 27.99 is approximately correct for
the dwarfs. It is important to note that the assignment of an 
individual galaxy to one of the two model branches may be less problematic 
than it first appears. As we have data for two (ideally more) SBF fields
per galaxy with a significant colour difference, we can decide on the basis
of colour slope. For instance, between the two fields of ESO384-016,
 we measured $\Delta \bar{M}_R=0.08(\pm0.08)$\,mag and 
$\Delta (B-R)_0=0.20(\pm0.06)$\,mag. This clearly indicates that ESO384-016
belongs on the blue branch. Due to smaller differences in colour 
the situation for ESO269-066, AM1339-445, 
and AM1343-452 is less clear. One could argue for each of them that 
it is actually a member of the blue branch and that the offset from the 
model values reflects the relative distance from the assumed group 
centre. However, this explanation appears highly unlikely as all 
three galaxies would by chance have to fall at the right place on 
the red branch.  

We prefer to interpret the results in Figure~\ref{fig7} {\em right panel} as a convincing 
demonstration how well theoretical models of mainly old, metal-poor stellar 
populations can predict the colour dependence of $\bar{M}_R$. The systematic 
offset between the dE data and the analytical expressions that 
describe the two branches (see above) is $-0.03$\,mag with a scatter of 
$\sigma=0.16$\,mag. This scatter accounts for genuine differences 
between models and observations as well as for the depth effect among the 
four dEs. Taking into account the zero point from the empirical data and 
the shape of the relation from the models we can formulate a semi-empirical 
calibration of the SBF method as distance indicator for dEs:

\begin{eqnarray}
\bar{M}_R = 6.09\cdot (B-R)_0-8.81
\end{eqnarray}

for the red, steeply rising branch in the colour range 
$1.10<(B-R)_0<1.50$, and 

\begin{eqnarray}
\bar{M}_R = 1.89\cdot [(B-R)_0-0.77]^2-1.26
\end{eqnarray}

for the blue branch in the range $0.80<(B-R)_0<1.35$. 
The decision which of the formulas is appropriate 
for a galaxy remains subject to the observed trend of $\bar{m}_R$ as a function of 
$(B-R)$ colour.  

We note that for fairly blue old galaxies, say with $(B-R)_0<1.0$, there is almost no
colour dependence of $\bar{M}_R$. For $(B-R)_0$ = 1.0 and 0.8, equation (3) gives
$\bar{M}_R$ = $-1.16$ and $-1.26$, respectively. This is in accord with
$\bar{M}_R \approx$ $-1.15$ adopted in our previous paper (JFB98) for the blueish
Sculptor group dwarf spheroidals.

\subsection{Resulting distances}
We are now in a position to derive individual distances for the dwarfs. First, by
using formulas (2) and (3) we have calculated the calibration constant for 
each SBF field, as listed 
in col.~[6] of Table~\ref{tbl-3}. In col.~[5] of that table we also give the 
``observed'' $\bar{M}_R$ based on $m-M$ = 27.99, as plotted in Figure~\ref{fig7}.
The errors in the calibrating $\bar{M}_R$ are simply the propagated colour errors
given in col.~[3] of Table~\ref{tbl-3}. ESO219-010 and 384-016 have been assigned to the 
blue branch (equation 3); the three other dwarfs to the main, red branch 
(equation 2). The errors in $\bar{M}_R$ are accordingly much smaller for the former
than the latter, as the red branch is quite steep. {\em Assuming no error in the 
models and in the assignment to them}, we now take any difference between 
$\bar{M}_R$(obs) and $\bar{M}_R$(calib) as a real difference in the distance 
between the object and the adopted mean group distance of $(m-M)$ = 27.99.
The individual distance moduli for the dwarf fields, listed in col.~[7] of Table~\ref{tbl-3},
follow then simply by combining $\bar{m}^0_R$ (col.~[4]) and $\bar{M}_R$(calib) (col.~[6]).
Finally, the corresponding distances for the fields are averaged for each galaxy (where 
two fields are available), giving the error-weighted mean distances in Mpc listed in 
col.~[8] of Table~\ref{tbl-3}. The distance errors vary between 3 and 9\% for the two 
blue branch dwarfs, and 13 to 19\% for the three red branch dwarfs.

The distance of ESO219-010, of 4.79 Mpc, is significantly larger than the distances
of the remaining four dwarfs (if assigned to the red branch in Figure~\ref{fig7},
the distance would even be 7.3 Mpc). While its Cen\,A group membership is not questioned,
this galaxy is likely lying outside the core region of the group, and it should be 
excluded for a calculation of the mean group distance. The four remaing dEs have
a mean distance of 4.00 Mpc (m$-$M = 28.01), which by chance exactly coincides
with the pre-assumed group distance inferred from the well-determined distances of 
NGC\,5128 and NGC\,5253. The small 1$\sigma$ distance dispersion of 
0.20 Mpc [or $\sigma (\rm m-M)$ = 0.10 mag] is not only 
a confirmation of the previous assumption that the Cen\,A group is well defined
and concentrated, but again shows that the SBF method for dwarf ellipticals works
quite well in this case.
%Finally, we combine the distances of NGC5128 and NGC5253 with the results for the 
%four dEs to determine an error-weighted new distance to the Cen\,A group 
%$(m-M)_{\mbox{Cen\,A}}=28.01(\pm0.04)$. Applying equal weight to the results
%yields $27.99(\pm0.04)$.

In Figure~\ref{fig9} we show the sky distribution of the 
Cen\,A group main members and the positions of the five dEs. We note that
ESO219-010 lies near the edge of this distribution, again suggesting that this 
is an outer member of the group. It is tempting to associate ESO384-016 with 
the galaxy pair NGC5253/5236, and to see ESO269-066 as a companion of AM1331-451. 
However, more distance and velocity information for the remaing galaxies is needed 
to unambiguously identify any subclumps, if present, in the Cen\,A group.

\section{SUMMARY AND CONCLUSIONS}
This is the second study on the surface brightness fluctuation 
method as useful distance indicator for low surface brightness 
dwarf elliptical galaxies. In a first paper (JFB98) we have 
applied the SBF method to dwarf spheroidal members of the nearby Sculptor
group. The Sculptor dwarfs (with one exception) are very faint
($M_B>-11$) and blue [$(B-R)_0^T<1.1$]. As with LG dwarf spheroidals (da\,Costa et 
al.~1996), many of them show signs of recent star formation.
This forced us to look, by means of stellar synthesis models provided
by Worthey (1994), for any variation of the calibrating $\bar{M}_R$
with age and metallicity of the underlying stellar population.
Very fortunately, {\em for the faint, blue, and metal-poor}\/ Sculptor
dwarfs this variation turned out to be minimal, and a preliminary
calibration constant of $\bar{M}_R \approx -1.15$ was adopted.

The present paper was intended to improve the SBF calibration and to extend
it to brighter magnitudes. 
We have carried out $B$ and $R$ CCD 
photometry of five dE galaxies, AM1339-445, AM1343-452, ESO219-010, 
ESO269-066, and ESO384-016, suspected members of the Centaurus\,A 
group. The SBF analysis of their $R$-band images revealed similar 
fluctuation magnitudes suggesting the same approximate distance to the 
Local Group (given their similar colour). The missing link for their 
association to the 
Cen\,A group is provided by the spectroscopic redshifts we measured 
for ESO269-066 and ESO384-016. Both dwarfs have heliocentric
velocities in good agreement with velocities found for main Cen\,A group members.

For the calibration of the measured apparent fluctuation magnitudes
we employed a mean distance modulus for the Cen\,A group of 3.96 Mpc
($m-M$ = 27.99 mag) based on
reliable (Cepheid, TRGB, PNLF) distances to NGC5128 and NGC5253. 
Again we had to study the dependence of the resulting absolute fluctuation 
magnitudes $\bar{M}_R$\/ on age and metallicity, i.e. in practice on
the $(B-R)_0$ colour of the underlying stellar population, by drawing 
on the on-line tool provided by Worthey (see footnote 1).

Here now comes the difference from the Sculptor dwarfs in our first paper.
The Cen\,A group dwarfs are more distant and hence on average brighter
than the Sculptor dwarfs. As a result, owing to a general colour-luminosity
relation (e.g., Ferguson 1994, Secker et al.~1997), the Cen\,A dEs are also
significantly redder, lying in the colour range $1.0<(B-R)_0<1.3$.
For these red colours the stellar population synthesis models show a strong
dependence of $\bar{M}_R$ on $(B-R)_0$. Moreover, two branches are populated
by the models in the $\bar{M}$-colour plane. One branch is more populated
by models based on Revised Yale Isochrones, the other by models based on
Padova Isochrones (Bertelli et al.~1994). We have argued that 
the Revised Yale Isochrones are  
unreliable for the purpose of predicting $\bar{M}_R$ magnitudes for old,
metal-poor, and sufficiently red stellar populations due to the lack of a 
full post red giant 
evolution. That the present dEs are dominated by old, metal-poor stars
is indeed confirmed by the absorption line spectra of 
ESO269-066 and ESO384-016.  

Aside from theoretical considerations, it was found that the fluctuation
magnitudes, when forced to a mean distance of 3.96 Mpc, are surprisingly
well matched by the model curves. An exception is ESO219-010 which seems
to be slightly more distant than the core of the Cen\,A group. We also found
that the assignment of a particular galaxy to one of the two branches
is fairly unambigous, especially if more than one subfield per galaxy is
analysed. In this way, by assuming that the scatter around the model curves,
given by equations (2) and (3)
is entirely due to a dispersion in the individual distances (depth effect),
we achieved SBF distances with an accuray of 3 to 9\% for the two dwarfs
belonging to the ``blue'' branch, and 13 to 19\% for the three dwarfs
belonging to the ``red'' branch. The distance dispersion of the dwarfs
(again excepting ESO219-010) is quite small, with a depth of ca.~0.5 Mpc,
confirming the isolation and compactness of the Cen\,A group.

As regards the application of the SBF method to dEs in general, the situation
for red dwarfs [$(B-R)_0>1.3$] is certainly much less favourable than for blue ones.
As future applications will tend to reach greater distances, the objects
will be brighter and hence redder. The slope of the $\bar{M}$-colour relation
in the $R$-band is very steep (the slope is 6.09, see equation 2), such that
accurate colours are prerequisite for a proper SBF analysis. In different 
photometric bands the $\bar{M}$-colour relation is somewhat less, but still 
considerably steep (e.g.~$4.5\pm0.25$ in $\bar{M}_I$ and $V-I$, see Tonry et al.~1997). 
Given the easy access and convenience of optical $R$-band imaging and the avoidance of 
fringing problems (as compared to the $I$-band) it is probably no great gain to change the
photometric band. As we have shown in this paper, by working with several subfields
per galaxy it should be feasible to achieve accurate SBF distances in the $R$-band
(with an error of less than, or of the order of 10\%) also for bright and red 
dwarf ellipticals. 
 
\acknowledgments

\begin{figure}
\centering\leavevmode
\epsfxsize=8.0cm
\epsfbox[17 143 486 696]{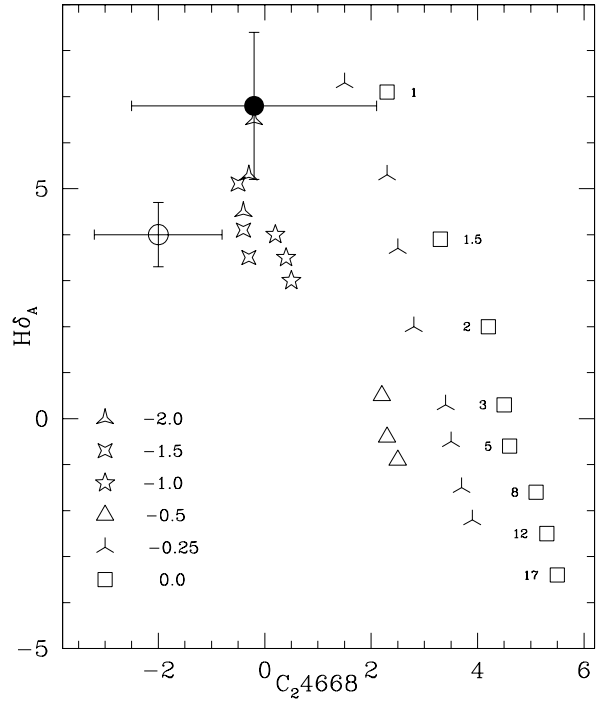}
\caption{The location of ESO269-066 (open circle) and ESO384-016 
(filled circle) in a H$\delta_{\mbox{A}}$ (age) versus C$_2$4668 (metallicity) grid derived 
from single-burst population models. Ages of 1, 1.5, 2, 3, 5, 8 ,12, and 17 Gyr 
are shown for [Fe/H]$>-0.25$. For more metal-poor systems only the three oldest 
models are shown. The uncertainty in the location of the dEs is considerably 
large, particularly for ESO384-016. Nevertheless, both dEs are likely to be 
metal-poor systems with a population dominated by old stars. 
\label{fig4}}
\end{figure}

\begin{figure}
\centering\leavevmode
\epsfxsize=10cm
\epsfbox[40 150 395 350]{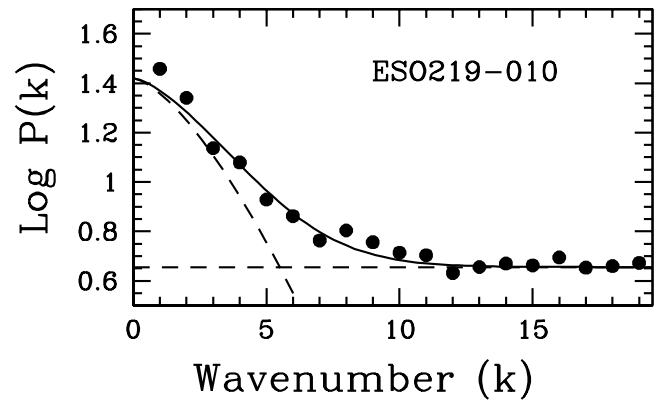}
\caption{Azimuthally averaged power spectra of the Fourier transform
of the processed $R$ image of ESO219-010. Due to many superimposed 
stars only one subimage could be analysed for this galaxy. The 
observations (filled circles) are well fitted by the sum of 
the PSF power spectrum and a constant (dashed lines).
\label{fig5}}
\end{figure}

\begin{figure*}[htb]
\centering\leavevmode
\SetRokickiEPSFSpecial
\HideDisplacementBoxes
\epsfxsize=15cm
\vspace{-2cm}
\BoxedEPSF{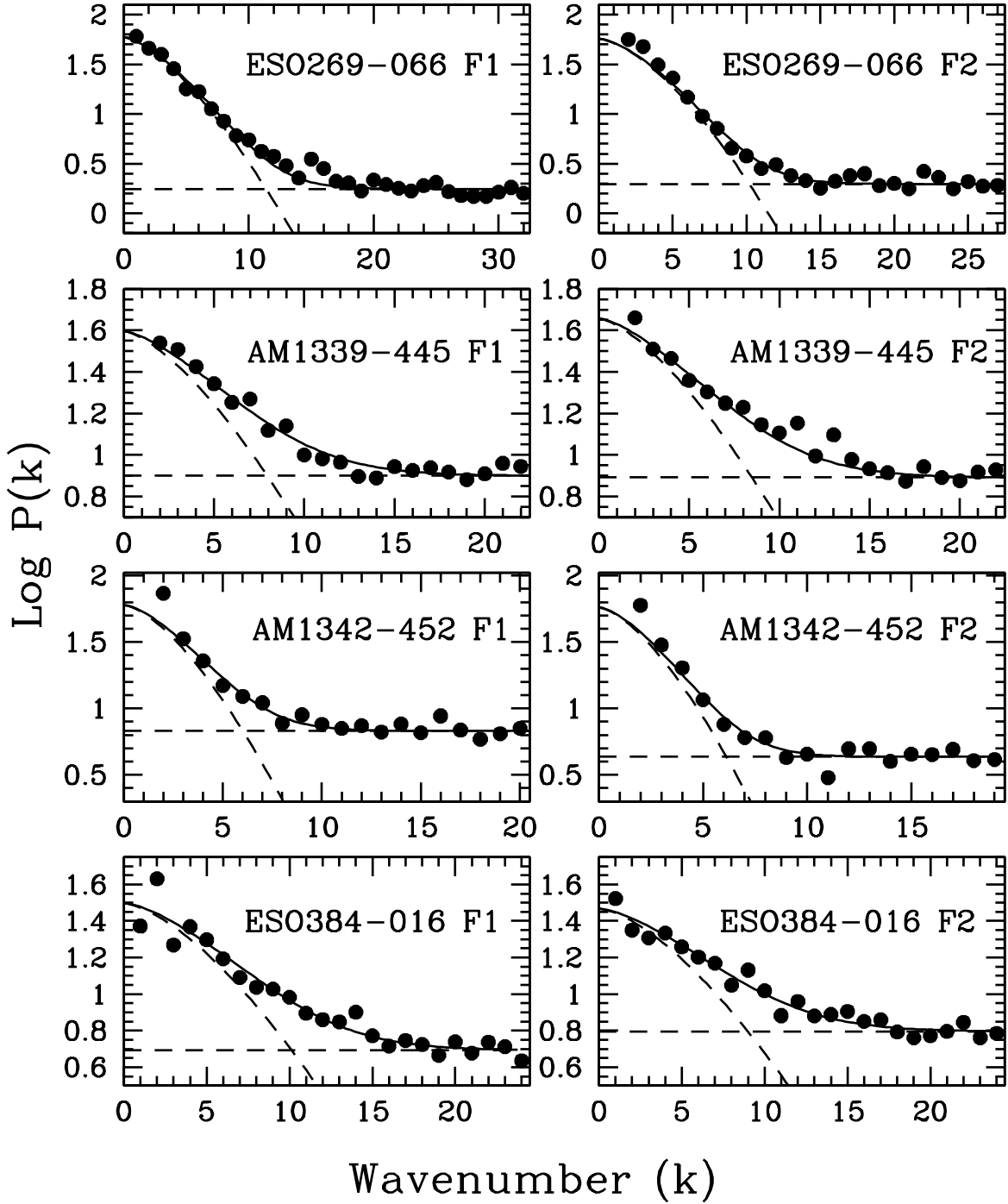}
\caption{Azimuthally averaged power spectra of the Fourier transform
of the processed $R$-band images of ESO269-066, AM1339-445, AM1342-452,
and ESO384-016. For each galaxy two non-overlapping
subimages were analysed (F1 and F2). The observations (filled circles) 
are well fitted by the sum of the PSF power spectrum and a constant 
(dashed lines).
\label{fig6}}
\end{figure*}

\begin{figure*}
\centering\leavevmode
\epsfxsize=17cm
\epsfbox[10 263 586 589]{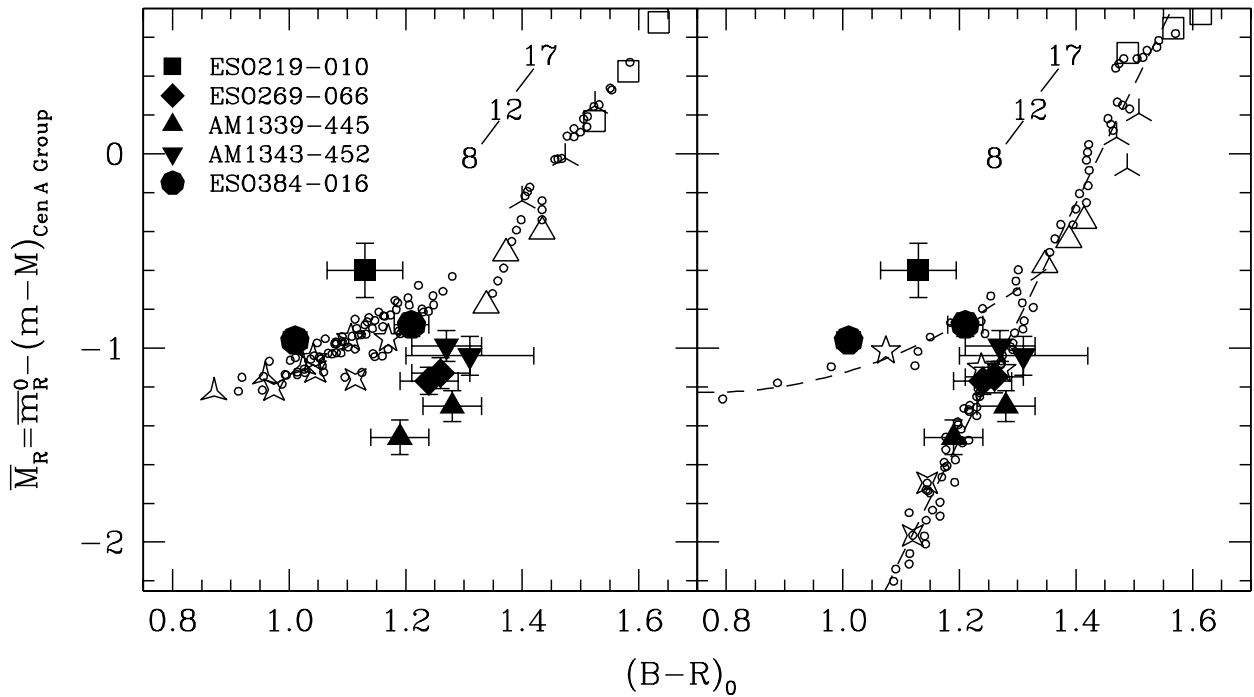}
\caption{Theoretical colour-fluctuation magnitude diagrams using Worthey's (1994) models.
The data in the left panel are based on the amalgamation of the stellar evolutionary 
isochrones of VandenBerg and collaborators and the Revised Yale Isochrones (Green et al.~1987) 
as described in Worthey (1994). For the right panel we employed the library of Bertelli et al.~(1994). 
The modelled pure stellar populations cover the \{age=8, 12, 17 Gyr\} $\times$ 
\{[Fe/H]=$-2.0$, $-1.9$, ..., $-1.0$, $-0.5$, $-0.25$, $0$\} parameter space (symbols as 
in Fig.~\ref{fig4}). In the case of the Bertelli isochrones, [Fe/H] was restricted to 
$\geq-1.7$ for 8 and 12\,Gyr, and $\geq-1.3$ for 17\,Gyr, respectively. Model values are 
also shown for the same stellar populations but polluted at the 10, 20 and 30\% 
level (in weight) by a 5 Gyr old second population with solar metallicity (open circles).
Altogether, the models indicate the theoretical locus of $\bar{M}_R$ magnitudes as a function of 
$(B-R)_0$ colours for a mainly old stellar population as typically observed in dwarf elliptical 
galaxies. Each data set exhibit two distinct branches. However, in each case only one is well 
defined while the other, though recognisable, is only sparsely populated. The dashed lines
in the right graph are the analytical forms which best fit the models. The two graphs are 
supplemented with the empirical results of the SBF analysis of nine fields in five dE 
galaxies. All galaxies are assumed to be Cen\,A group members with $(m - M)_0$ = 27.99.
\label{fig7}}
\end{figure*}

\begin{figure*}
\centering\leavevmode
\epsfxsize=17cm
\epsfbox{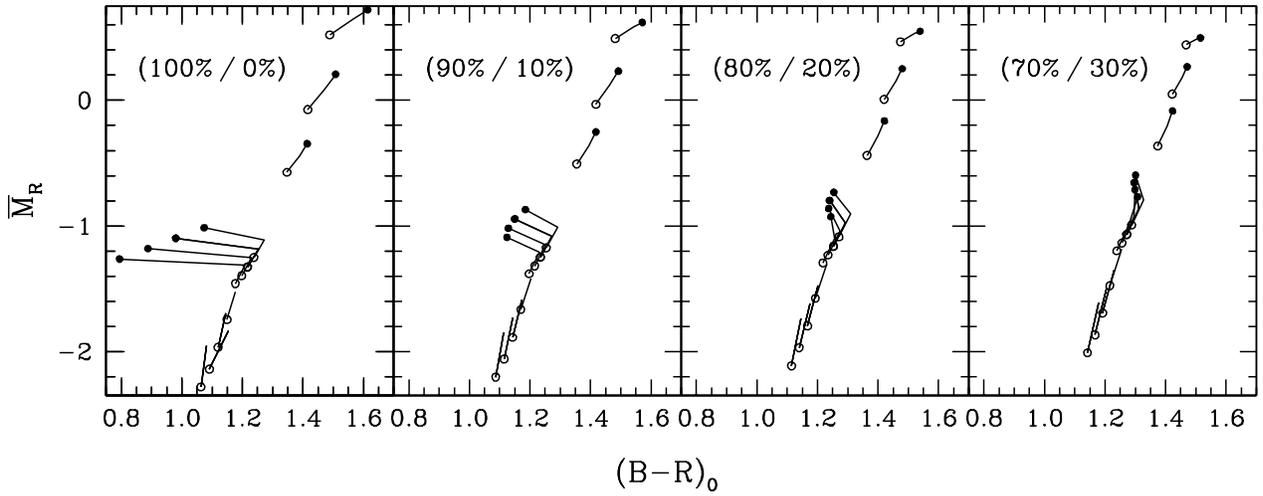}
\caption{The colour-fluctuation magnitude relation based on the library of Bertelli 
et al.~(1994) as shown in Fig.~\ref{fig7} but separated according to the increasing 
fraction of a 5 Gyr old second stellar population (from left to right). The tracks 
in each panel correspond to different metallicities, from [Fe/H]=$-1.7$ (bottom)
to [Fe/H]=$0.0$ (top). Each track starts at 8\,Gyr (open circle) and ends either at 
12\,Gyr, if $-1.7\leq$[Fe/H]$\leq-1.4$, or at 17\,Gyr if $-1.4\leq$[Fe/H] (filled circle).
The location of the model values clearly indicate that only the very old, metal-poor
stellar populations contribute to the blue branch. 
\label{fig8}}
\end{figure*}

\begin{figure*}
\centering\leavevmode
\epsfxsize=13cm
\epsfbox{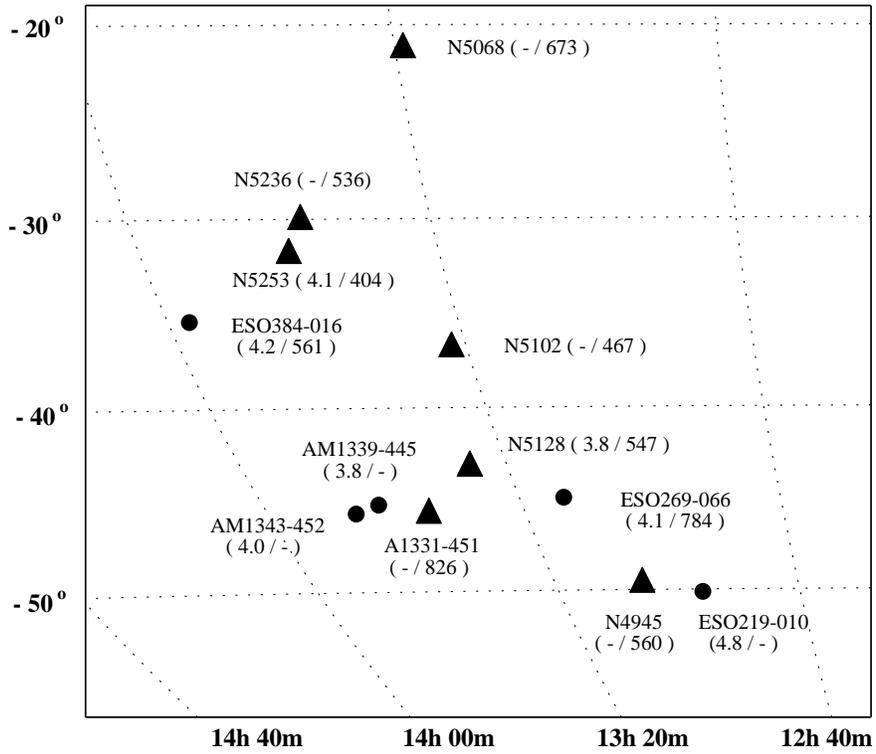}
\caption{Distribution in R.A.~(J2000.0) and Decl.~(J2000.0) of the 
Centaurus\,A group main members (filled triangles) and the new dEs 
(filled circles). Distances in Mpc and heliocentric velocities in \kms  
are indicated in parentheses if available. \label{fig9}}
\end{figure*}

We thank Lewis Jones for interesting discussions and for providing Figure~\ref{fig4}. 
H.J. and B.B. are grateful to the {\em Swiss National Science Foundation} for financial 
support. H.J. thanks further the Freiwillige Akademische Gesellschaft der Universit\"at 
Basel and the Janggen-P\"ohn Stiftung for their partial financial support.
This project made extensively use of Worthey's World Wide Web model interpolation 
engine and made use of the NASA/IPAC Extragalactic Database (NED), which 
is operated by the Jet Propulsion Laboratory, Caltech, under contract with 
the National Aeronautics and Space Administration.

\newpage

%%%%%%%%%%%%%%%%%%%%%%%%%%%%%%%
% Figures
%%%%%%%%%%%%%%%%%%%%%%%%%%%%%%%

\clearpage

\clearpage

\clearpage

\clearpage

\clearpage

\clearpage

\clearpage

\clearpage

\clearpage

%%%%%%%%%%%%%%%%%%%%%%%%%%%%%%%
% Tables
%%%%%%%%%%%%%%%%%%%%%%%%%%%%%%%

\begin{deluxetable}{cccccccccc}
\scriptsize
\tablecaption{Fundamental Parameters of Five Dwarf Elliptical Galaxies in the Cen\,A Group \label{tbl-1}}
\tablewidth{0pt}
\tablehead{
\colhead{}       & \colhead{}    &\colhead{R.A.}     & \colhead{Decl.}  &\colhead{$B_T$} &\colhead{$R_T$} & 
\colhead{$r_{eff,R}$} & \colhead{$\langle \mu\rangle_{eff,R}$} & \colhead{A$_B$}& \colhead{$(B-R)_0^T$} \\
\colhead{Name} &\colhead{Type} & \colhead{(J2000.0)}   & \colhead{(J2000.0)} & \colhead{(mag)} & \colhead{(mag)} & 
\colhead{(arcsec)} & \colhead{(mag\,arcsec$^{-2}$)}& \colhead{(mag)}      & \colhead{(mag)} \\
\colhead{(1)}    & \colhead{(2)} & \colhead{(3)}  & \colhead{(4)}   & \colhead{(5)}  & 
\colhead{(6)}    & \colhead{(7)} & \colhead{(8)}  & \colhead{(9)}   & \colhead{(10)} }      
\startdata  
ESO219-010& dE,N     \hfill &12h56m09.6s & $-$50d08m38s & 16.42 & 14.73 & 17.4 & 22.92 & 0.96  & 1.32 \\    
ESO269-066&dE,N:     \hfill &13h13m09.5s & $-$44d52m56s & 14.59 & 12.96 & 39.7 & 22.95 & 0.40  & 1.48 \\    
AM1339-445&dE        \hfill &13h42m05.8s & $-$45d12m21s & 16.32 & 14.76 & 23.8 & 23.63 & 0.48  & 1.38 \\     
AM1343-452&dE        \hfill &13h46m17.8s & $-$45d41m05s & 17.57 & 16.02 & 14.7 & 23.85 & 0.52  & 1.35 \\     
ESO384-016&dS0/Im    \hfill &13h57m01.2s & $-$35d19m59s & 15.11 & 13.90 & 19.2 & 22.31 & 0.32  & 1.09 \\     
\enddata
\end{deluxetable}

\clearpage

\begin{deluxetable}{ccccccccccc}
\scriptsize
\tablecaption{Parameters of the SBF Analysis \label{tbl-2}}
\tablewidth{0pt}
\tablehead{
\colhead{} &\colhead{$m_1$} & \colhead{Exp. Time} & \colhead{FWHM} & \colhead{$\bar{g}$}  &\colhead{$s(\Delta s)$} 
&\colhead{$P_0(\Delta P_0)$} & \colhead{$P_1(\Delta P_1)$} &\colhead{$\bar{m}_{R}(\Delta \bar{m})$} & \colhead{A$_R(\Delta$A$_R)$}  
&\colhead{$\bar{m}_R^0(\Delta \bar{m})$} \\
\colhead{Name}       & \colhead{(mag)}              & \colhead{(s)}   & \colhead{(arcsec)}      & \colhead{(ADU)}          & \colhead{(ADU)}  & 
\colhead{(ADU)}       & \colhead{(ADU)} & \colhead{(mag)}   & \colhead{(mag)}      & \colhead{(mag)}   \\
\colhead{(1)}    & \colhead{(2)}              & \colhead{(3)}   & \colhead{(4)}      & \colhead{(5)}          & \colhead{(6)} &
\colhead{(7)}    & \colhead{(8)} & \colhead{(9)}      & \colhead{(10)}          & \colhead{(11)} 
}     
\startdata
ESO219-010    \dotfill& 24.58 & 5$\times$ 600 & 1.4 & 242 & 2572(3)& 26.3(2.7) & 4.5(0.05) &  27.98(0.10) & 0.59(0.09) & 27.39(0.14)\\
						                  			    
ESO269-066 F1 \dotfill& 24.56 & 5$\times$ 600 & 1.3 & 473 & 2120(2)& 59.4(3.3) & 1.8(0.08) &  27.07(0.06) & 0.25(0.04) & 26.82(0.07)\\
           F2 \dotfill&       &               &     & 447 &        & 56.9(3.8) & 2.0(0.08) &  27.11(0.07) &            & 26.86(0.08)\\
						                 			    
AM1339-445 F1 \dotfill& 24.03 & 4$\times$ 600 & 1.2 & 223 & 3566(7)& 39.4(2.4) & 7.9(0.10) &  26.99(0.06) & 0.30(0.05) & 26.69(0.08)\\
           F2 \dotfill&       &               &     & 204 &        & 45.4(2.9) & 7.8(0.14) &  26.83(0.07) &            & 26.53(0.09)\\
						                 			    
AM1343-452 F1 \dotfill& 24.58 & 5$\times$ 720 & 1.4 & 162 & 2896(4)& 60.2(5.0) & 6.8(0.13) &  27.27(0.09) & 0.32(0.05) & 26.95(0.10)\\
           F2 \dotfill&       &               &     & 226 &        & 57.6(3.3) & 4.3(0.09) &  27.32(0.06) &            & 27.00(0.08)\\
						                  			    
ESO384-016 F1 \dotfill & 24.03 & 5$\times$ 600 & 1.1 & 428 & 3290(6)& 31.5(1.0) & 4.9(0.07) & 27.23(0.04) & 0.20(0.03) & 27.03(0.05)\\
           F2 \dotfill &       &               &     & 205 &        & 29.4(1.2) & 6.2(0.08) & 27.31(0.05) &            & 27.11(0.06)\\
\enddata
\end{deluxetable}

\clearpage

\begin{deluxetable}{cccccccc}
\tablecaption{Local Colours, Fluctuation Magnitudes, Luminosities, and Distances\label{tbl-3}}
\tablewidth{0pt}
\tablewidth{0pt}
\tablehead{
\colhead{}     &\colhead{}       & \colhead{$(B-R)_0$} & \colhead{$\bar{m}_R^0$} & 
\colhead{$\bar{M}_R$(obs)}  &\colhead{$\bar{M}_R$(calib)} &\colhead{(m-M)$_0$} &\colhead{D}    \\
\colhead{Name} & \colhead{Field} & \colhead{(mag)}     & \colhead{(mag)}                         & 
\colhead{(mag)}             & \colhead{(mag)}              &\colhead{(mag)}  &\colhead{(Mpc)}\\
\colhead{(1)}    & \colhead{(2)}              & \colhead{(3)}   & \colhead{(4)}      & 
\colhead{(5)}          & \colhead{(6)} & \colhead{(7)} & \colhead{(8)} 
}     
\startdata
ESO219-010 \dotfill& F1 & 1.13(0.08) & 27.39(0.14) & $-$0.60 & $-$1.01(0.11) 
& 28.40(0.18) & 4.79(0.43)\\
                                               & F1 & 1.24(0.06) & 26.82(0.07) & $-$1.17 & $-$1.26(0.36) 
& 28.08(0.37) & \\
\raisebox{1.5ex}[0cm][0cm]{ESO269-066\dotfill} & F2 & 1.26(0.06) & 26.86(0.08) & $-$1.13 & $-$1.14(0.36) 
& 28.00(0.37) & \raisebox{1.5ex}[0cm][0cm]{4.05(0.53)} \\
                                               & F1 & 1.28(0.06) & 26.69(0.08) & $-$1.30 & $-$1.01(0.36) 
& 27.70(0.37) & \\
\raisebox{1.5ex}[0cm][0cm]{AM1339-445\dotfill} & F2 & 1.19(0.06) & 26.53(0.09) & $-$1.46 & $-$1.56(0.36) 
& 28.09(0.37) &  \raisebox{1.5ex}[0cm][0cm]{3.75(0.49)}\\
                                               & F1 & 1.31(0.12) & 26.95(0.10) & $-$1.04 & $-$0.83(0.73) 
& 27.78(0.74) & \\
\raisebox{1.5ex}[0cm][0cm]{AM1343-452\dotfill} & F2 & 1.27(0.07) & 27.00(0.08) & $-$0.99 & $-$1.08(0.42) 
& 28.08(0.43) & \raisebox{1.5ex}[0cm][0cm]{3.97(0.74)}\\
                                               & F1 & 1.01(0.04) & 27.03(0.05) & $-$0.96 & $-$1.15(0.04) 
& 28.18(0.06) & \\
\raisebox{1.5ex}[0cm][0cm]{ESO384-016\dotfill} & F2 & 1.21(0.05) & 27.11(0.06) & $-$0.88 & $-$0.89(0.08) 
& 28.00(0.10) & \raisebox{1.5ex}[0cm][0cm]{4.23(0.11)}\\
\enddata
\end{deluxetable}

\clearpage

\begin{deluxetable}{ccccc}
%\scriptsize
\tablecaption{Distances and Velocities of Two Main Cen\,A Group Galaxies \label{tbl-4}}
\tablewidth{0pt}
\tablehead{
\colhead{}   &\colhead{$v_\odot$} &\colhead{(m-M)$_0$} & \colhead{}  &\colhead{}     \\
\colhead{Name}   & \colhead{(km\,s$^{-1}$)}   & \colhead{(mag)}     & \colhead{Estimator\tablenotemark{a}} & \colhead{ref}  \\
\colhead{(1)}  & \colhead{(2)}     &\colhead{(3)} & \colhead{(4)}& \colhead{(5)}}    
\startdata
NGC5128 \dotfill & 562      & 27.98($\pm 0.15$)& TRGB\hfill   & Harris et al.~1999 \hfill   \\
        \dotfill & \dotfill & 27.97($\pm 0.14$)& PNLF\hfill   & Hui et al.~1993 \tablenotemark{b}\hfill      \\
        \dotfill & \dotfill & 27.78($\pm 0.10$)& SBF\hfill    & Tonry \& Schechter 1990 \tablenotemark{b}\hfill   \\
NGC5253 \dotfill & 404      & 28.08($\pm 0.10$)& Cepheids\hfill & Saha et al.~1995 \hfill         \\
\enddata
\tablenotetext{a}{TRGB: Tip of the Red Giant Branch magnitude, PNLF: Planetary Nebulae Luminosity 
Function, SBF: Surface Brightness Fluctuations.} 
\tablenotetext{b}{Adjusted to a Local Group distance scale set by (m -- M)$_0$(M31) = 24.5.} 
\end{deluxetable}

\end{document}